\documentclass[11pt]{article}

\usepackage[letterpaper, margin=1in]{geometry}
\usepackage[T1]{fontenc}
\usepackage{mathpazo}
\usepackage{microtype}
\usepackage{amsmath,amssymb}
\usepackage{graphicx}
\usepackage{xcolor}

\definecolor{linkblue}{RGB}{30, 80, 180}
\definecolor{linkgreen}{RGB}{30, 120, 60}
\definecolor{linkred}{HTML}{AE0000}

\usepackage{booktabs}
\usepackage{enumitem}
\usepackage[numbers,sort&compress]{natbib}
\usepackage{array}        
\usepackage{subfiles}     
\usepackage{hyperref}
\hypersetup{
  colorlinks=true,
  urlcolor=linkblue,
  linkcolor=linkgreen,
  citecolor=linkgreen,
  filecolor=linkred,
}

\graphicspath{
  {figures/}%
  {../Figures/}%
  {/Users/bozprog/Library/Mobile Documents/com~apple~CloudDocs/FIGURES/}%
}

\newcommand{\OAE}{Open \AE thernet}

\title{The Bilateral Efficiency of Ethernet:\\
Recalibrating Metcalfe and Boggs After Fifty Years}

\author{Paul Borrill\\
\textit{D{\AE}D{\AE}LUS --- Open \AE thernet Project, Asilomar, CA}\\
\texttt{paul@daedaelus.com}}

\date{v0.6 --- November 24, 2026\\[4pt]
\small 50th Anniversary of Metcalfe \& Boggs,
\textit{Comm.\ ACM} 19(7), July 1976}

\begin{document}
\maketitle

\begin{abstract}
In July 1976, Metcalfe and Boggs published their foundational paper on
Ethernet in \textit{Communications of the ACM}. Their efficiency
model---$E = (P/C)/(P/C + W\cdot T)$---measures the fraction of Ether time
carrying good \emph{forward} packets under contention. For fifty years this
model has framed how the community thinks about Ethernet performance. We
argue that the model, while correct for its purpose, is silent on the
question that matters for modern intra-rack interconnect: \emph{bilateral
transaction efficiency}---the fraction of link time that produces committed
agreements between sender and receiver. We show that Metcalfe and Boggs
themselves planted the seed of bilateral transactions in their EFTP
``end-dally'' protocol (Section~7.2.2), and that the deeper historical
anchor is older still: Abramson's Alohanet carried positive acknowledgments
at the link layer---a bilateral mechanism Metcalfe consciously \emph{removed}
in 1973 to obtain Ethernet's simple, ACK-free packet format. The result is a
fifty-year \emph{bilateral zigzag}: Aloha (bilateral) $\to$ Ethernet
(unilateral) $\to$ EFTP end-dally (bilateral) $\to$ TCP
(unilateral-with-bilateral-above). We formalize bilateral efficiency,
connect it to the back-to-back Shannon channel with Perfect Information
Feedback, and---scoping the claim explicitly to intra-rack distances of one
meter or less---describe how the \emph{Open \AE thernet} link recovers
mutual knowledge at the link layer. The correction to Table~1 is not a
different set of numbers. It is a different \emph{question}. A new appendix places the 2025--2026
return of link-level retry (UEC, OCP ESUN, UALink) in this history and shows
that it restores half of the end-dally: the sender learns of delivery, the
receiver never learns that the sender knows. It opens from Ron Crane's
observation that produced Fast Ethernet: carried to 100~Gb/s, the same
arithmetic makes the minimum frame about one meter long.
\end{abstract}

\section{Introduction}
\label{sec:intro}

On its fiftieth anniversary, Metcalfe and Boggs' 1976 paper
\cite{metcalfe1976ethernet} remains one of the most influential publications
in computing. It introduced not merely a technology but a design philosophy:
distributed control, passive shared media, statistical arbitration, and---
decisively---the separation of packet transport from reliable delivery. As
Robert Garner has documented from the primary sources, Metcalfe
\emph{consciously removed} link-level acknowledgments during Ethernet's
formulation in early 1973: the ARPANET~IMP carried a 1-bit Ack field,
Alohanet carried positive acknowledgments, and Metcalfe diverged from both
to keep the packet format simple \cite{garner2026personal}. That removal,
Garner argues, ``is what enabled Ethernet to be as successful as it has
been''---a designer at a small adapter company never had to implement ACK
buffers, timeout state machines, or retransmission queues. We take this
historical claim as correct and as the starting point, not the target, of
the argument.

The efficiency model that produces the paper's celebrated Table~1 measures
one thing precisely: what fraction of Ether time carries good forward
packets. It says nothing about whether those packets arrive, whether the
receiver processes them, or whether sender and receiver \emph{agree} a
transfer completed. This was the right scope for 1976, when the open
question was whether a shared coax could carry useful traffic under
contention. The scope limitation has since been mistaken for the complete
answer.

We argue that the right question for 2026---in the specific regime of
intra-rack interconnect at 800~Gb/s and beyond---is not ``what fraction of
link time carries good forward packets?'' but ``what fraction of link time
produces \emph{committed bilateral agreements}?'' The paper develops this in
five steps: (1)~the bilateral zigzag through Ethernet's own history;
(2)~the end-dally as a bilateral transaction Metcalfe built and then
confirmed in personal communication; (3)~the distinction between forward
efficiency and bilateral efficiency, with the two timing bounds separated so
they cannot be read as contradictory; (4)~the back-to-back Shannon channel
with Perfect Information Feedback, including the pipelining that keeps its
marginal cost near zero; and (5)~an explicit fault model and scope, so the
claim can be evaluated---or falsified---at the engineering level.

A note on what this paper does \emph{not} do. It carries no quantum-mechanical
argument. Earlier drafts invoked the Two-State Vector Formalism as an analogy
for two-boundary transactions; careful readers (Robert Garner and the Mulligan Stew Gang)
independently judged that the analogy cost more than it bought for
an engineering audience, and they were right. The bilateral argument stands
on Bill Lynch's alternating-bit protocol and on Metcalfe and Boggs' own text,
with no appeal to physics beyond the propagation delay of a wire.

\section{A Note on the Name: Open \AE thernet}
\label{sec:name}

Garner has pressed one objection longer and harder than any other: this is
not Ethernet, and it should not be called Ethernet. ``Ethernet'' is a brand
stewarded for five decades through the IEEE~802 process; its defining
commitment is precisely the \emph{absence} of link-level acknowledgment that
this work proposes to restore. To attach the name to a link layer that
reintroduces bilateral commitment is, in his words, to hitch one's horse to
the queen of the parade. The objection is fair, and we concede it.

We therefore name the architecture \textbf{Open \AE thernet}. The ligature is
not decoration and it is not ``Atomic-Ethernet'' smuggling the brand back in.
It is the Greek privative \emph{alpha}---the negating prefix of
\emph{a}-cyclic, \emph{a}-moral, \emph{a}-temporal, \emph{a}-byss---fused with
the luminiferous \emph{aether} that nineteenth-century physics invoked to
carry light and that twentieth-century physics abolished. \emph{\AE thernet}
thus names, by construction, a thing that is \emph{not} the classical
ether-net: a link layer whose contract is bilateral commitment rather than
best-effort forwarding, while openly honoring the lineage from which it
descends. The privative reading also rhymes with the engineering content:
\AE thernet is the network that removes the unilateral
timeout-and-retry assumption (``\emph{a}-timeout'') rather than the network
that adds yet another forward-error layer. We adopt the name as a genuine
concession to Garner's stewardship of the Ethernet brand, and we record our
debt to him for forcing the question.

\section{The Bilateral Zigzag: Aloha to Ethernet to EFTP}
\label{sec:zigzag}

The cleaner way to tell Ethernet's history is not as a single design choice
but as an oscillation between bilateral and unilateral mechanisms at
different layers across fifty years.

\begin{description}[style=nextline,leftmargin=1.5em,nosep]
\item[Alohanet (bilateral).] Abramson's system used positive
  acknowledgments. As Garner puts it, ``Alohanet worked fine enough without
  carrier or collision detection. Positive Acks provided bilateral
  agreement.'' The bilateral mechanism lived \emph{at the link}.
\item[Ethernet (unilateral).] Metcalfe removed the Acks in 1973. The link
  became best-effort and ACK-free; bilateral agreement was pushed upward.
  This is the divergence Garner documents and credits for Ethernet's
  success.
\item[EFTP end-dally (bilateral, one layer up).] In the very same 1976
  paper, Section~7.2.2 reintroduces a bilateral handshake at the
  file-transfer layer (Section~\ref{sec:end-dally}). Bilateral agreement
  reappears---just not at the link.
\item[TCP (unilateral-with-bilateral-above).] PUP~$\to$~BSP~$\to$~TCP
  (the CSL-79-10 ``blue and white'' lineage Garner supplied) carries
  bilateral mechanisms---BSP's positive acknowledgments with dallying---above
  a best-effort link. The zigzag persists.
\end{description}

We previously framed CSMA/CD collision detection itself as the ``original
bilateral primitive.'' Garner objected that this is ``stretching it''---
collisions are an undesirable side effect of imperfect carrier sense, a
``hack on top of a hack,'' never intended to be reliable---and he is correct.
We withdraw that claim. The honest historical anchor for bilateral agreement
is not collision detection; it is Alohanet's positive Acks, the mechanism
Metcalfe removed. \AE thernet's proposition is narrow and historical: restore
\emph{at the link} the bilateral commitment that Aloha had and Ethernet
dropped---now that the link technology (sub-meter, 800~Gb/s) and the workload
(coordination-bound AI clusters) make the dropping expensive. Appendix~\ref{app:ackreturns} records a fifth
swing, under way since 2025, in which industry consortia are restoring
link-level acknowledgment, and asks how much of the bilateral commitment it
restores.

\section{The End-Dally: A Bilateral Transaction in 1976}
\label{sec:end-dally}

Section~7.2.2 of Metcalfe and Boggs describes the EFTP ``End'' protocol for
completing a \emph{file} transfer. (It is per-file, not per-packet---a
correction we owe to Garner.) The sender transmits an END packet; the
receiver responds with a matching ENDREPLY and then \emph{dallies} for ``some
reasonably long period of time (10 seconds)''; the sender transmits an
echoing ENDREPLY and ``is free to go off with the assurance that the file has
been transferred successfully. The dallying receiver then gets the echoed
endreply and it too goes off assured'' \cite{metcalfe1976ethernet}.

The word \emph{assured} appears twice---once for the sender, once for the
receiver. Metcalfe and Boggs are reaching for \textbf{mutual knowledge}: both
parties know that the transfer succeeded, and the echoed endreply also tells
the receiver that the sender knows. The
10-second dally is a timeout-based approximation of what a bilateral link
transaction achieves deterministically.

In March 2026 Metcalfe confirmed this reading: ``As I recall, the end-dally
was intended to speed things up. Like the NAK does in general''
\cite{metcalfe2026personal}. The comparison is revealing. NAK-based protocols
are faster than ACK-timeout protocols precisely because
\emph{backward-flowing} information---the receiver reporting an outcome---lets
the system converge sooner than hopeful silence. Systems get \emph{more}
efficient, not less, when backward information flows freely. Yet the
efficiency model published in the same paper measures only the forward
channel. The metric and the protocol design were optimizing for different
things.

\section{What the Efficiency Model Measures---and What It Does Not}
\label{sec:efficiency}

Metcalfe and Boggs' efficiency, for $Q$ continuously queued stations on a
shared medium, is
\begin{equation}
\label{eq:E-forward}
E = \frac{P/C}{P/C + W\cdot T},
\end{equation}
with $P$ the packet size in bits, $C$ the channel capacity, $T$ the slot
(one round-trip propagation delay), and $W=(1-A)/A$ the mean contention
slots, $A=(1-1/Q)^{Q-1}$. The model is elegant and correct for its purpose,
and---as Garner notes and as Metcalfe's own 1973 thesis makes explicit via
the SF-10 effective-capacity formula
$\text{EffCap}=(S/P)\cdot\frac{1}{1+CT/P}\cdot(1-L)\cdot C$
\cite{metcalfe1973thesis}---Equation~\eqref{eq:E-forward} applies only to the
shared-media/cable Ethernet. On a modern switched, full-duplex link there is
no contention and $E\approx 1$ trivially. If the original efficiency model
does not even apply to modern Ethernet, the question of what should replace
it is wide open.

The model is silent on: receiver processing (the gap between physical
delivery and semantic agreement); bilateral completeness (whether both sides
agree); end-dally overhead; retransmission cost (Garner concedes this is a
real cost); and semantic correctness (which he also concedes). It has no term
for backward information---no variable for the receiver's contribution---even
though, as Metcalfe observed, the backward NAK is what speeds the system up.

\section{Bilateral Efficiency, and Two Timing Bounds That Are Not in Conflict}
\label{sec:bilateral}

Define \textbf{bilateral efficiency}
\begin{equation}
\label{eq:E-bilateral}
E_B = \frac{\text{committed transactions}}{\text{total link-seconds}},
\end{equation}
where a transaction counts as committed only when it is both
\emph{committed} and \emph{complete}: the deciding end has recorded the
outcome in state that survives the faults of the link's model (the commit
point), and the outcome has reached the other end, so that both sides know
it. A third condition, \emph{closure}, governs reuse rather than counting:
the link returns to rest before it carries the next exchange. Mutual
knowledge here means that each side knows the outcome; it is not common
knowledge, which no finite exchange over an unreliable channel attains
(Appendix~\ref{app:ackreturns}). A transaction counted at the link is not
thereby counted at the layer above, which must meet the same conditions
with its own peer (Appendix~\ref{app:ackreturns}). For the original shared-media Ethernet, $E_B \ll E$ once the ACK
round-trips, the three-phase end-dally, and retransmissions are charged. For
modern switched links, $E\approx 1$ while $E_B$ may be arbitrarily poor: an
RDMA ``completion'' reports success at physical delivery while the
application may have no semantic agreement.

Two timing quantities have been read as contradictory; they are not, because
they live at different scopes. We label them explicitly.

\paragraph{The per-commitment lower bound, $T \ge 2\tau$.}
A \emph{single} bilateral commitment cannot complete faster than one
round-trip propagation $2\tau$: the confirmation must physically return
before either side may declare the transaction committed. This is a hard
floor, the bilateral analog of Metcalfe's collision-detection slot
constraint, and \AE thernet does not evade it.

\paragraph{The marginal cost over forward-only, $\Delta T_{\text{commit}}\approx 0$.}
Once the protocol is \emph{pipelined} (Section~\ref{sec:shannon}), the
\emph{extra} wall-clock cost of bilateral confirmation, over what a
forward-only protocol would have spent moving the same payload across the
same link, is structurally near zero: the SACK return rides the otherwise-idle
backward path concurrently with the next forward frame.

Both hold together: a single commitment costs at least one
round-trip, \emph{and} the marginal cost of making delivery bilateral rather
than forward-only is $\approx 0$. Equation~\eqref{eq:E-forward} has no term
for either, because it has no term for backward information at all.

\section{Back-to-Back Shannon Channels, PIF, and Pipelining}
\label{sec:shannon}

The Open \AE thernet link is two concurrent Shannon channels---forward and
backward---with slice-by-slice feedback, rather than one forward channel with
ACKs bolted on. It defines four progressive levels of knowledge depth, each
carried by a 2-bit Slice ACK (SACK):

\begin{description}[nosep,leftmargin=2em,style=nextline]
\item[SACK 00 --- detection] first 8-byte slice seen without error; working
  cable and SerDes confirmed.
\item[SACK 01 --- capture] 16 bytes captured; alignment and expected header
  fields verified.
\item[SACK 10 --- recognition] 32 bytes received; protocol/message type
  determined; buffer and ring descriptors confirmed available.
\item[SACK 11 --- commitment] full 64 bytes received and matched to a
  legitimate frame layout, FCS valid; ready for the ring buffer. No further
  layer-2 repair needed.
\end{description}

This is Abramson's \emph{Perfect Information Feedback} (PIF)
\cite{abramson1973packet} realized at the physical layer: the sender witnesses
its own transmission returning, slice by slice. Crucially, each rung is a
\emph{witness} to the prior rung---SACK~11 is meaningful only if SACK~10 was,
and so on down---so commitment is an outcome both endpoints reach, each from
witnesses it has received, not a guess the sender makes after the fact. Why a structural transform rather
than plain echo-and-compare, and why this is trustworthy with no
error-correcting code on the SACK bits, is exactly the question Garner
pressed; we summarize the answer in Section~\ref{sec:faultmodel} and prove it
in Appendix~\ref{app:faultmodel} (the OAE fault model).

\paragraph{Where PIF lives (it is not asking applications to change).}
Garner reasonably notes that existing network applications do not run a
PIF-style ``compare what was sent to what was echoed'' loop---and they do not
need to. PIF is at the link layer (SerDes/PHY/NIC), not the application. An
application calls \texttt{send()} and \texttt{recv()} exactly as today and
never sees the witness round-trip. What changes is only the \emph{completion
semantic} the link exposes: ``this transaction committed bilaterally between
NICs in $N$ microseconds'' rather than ``some bytes left the door; verify
later.'' This is the abstraction-boundary question from the outset of this
correspondence---\AE thernet sits intentionally below Ethernet's traditional
abstraction boundary. We are aligned on the layering; the open disagreement is
whether the layering is worth doing.

\paragraph{Pipelining (why one outstanding transaction does not mean
stop-and-wait).} Garner's concern is well founded: with a single outstanding
transaction and no pipelining, throughput degrades to stop-and-wait, and on a
ten-meter link utilization would indeed collapse to a few percent. \AE thernet
windows the link the way TCP windows a connection: consecutive frames are in
flight on the forward path while their SACKs return on the backward path; the
sender's only blocking condition is the outstanding-bilateral-witnesses
window, not per-frame round-tripping. With windowing, throughput is bounded by
forward bandwidth and the marginal confirmation cost is the
$\Delta T_{\text{commit}}\approx 0$ of Section~\ref{sec:bilateral}. Within the
scope of Section~\ref{sec:scope-note}---sub-meter links where a single
minimum-size frame at 800~Gb/s already fills the wire---the pipe is full with
even one frame outstanding, which is the ``the snake fills the building''
observation Metcalfe offered the author at dinner.

The bilateral efficiency of this channel is then
\begin{equation}
\label{eq:E-bilateral-oae}
E_B^{\text{\AE}} =
\frac{N_{\text{committed}}}{N_{\text{attempted}}}
\cdot
\frac{P_{\text{eff}}}{P_{\text{eff}} + \Delta T_{\text{commit}}},
\qquad \Delta T_{\text{commit}}\approx 0,
\end{equation}
with the $T\ge 2\tau$ floor of Section~\ref{sec:bilateral} applying to each
individual commitment.

\section{Scope, and the Fault Model in Brief}
\label{sec:scope-note}\label{sec:faultmodel}

\paragraph{Scope (stated up front).} Open \AE thernet targets links of
\textbf{one meter or less}---intra-rack, shrinking to sub-centimeter at the
chiplet/module boundary---and a cellular mesh topology whose spanning-tree
multiplicity (matrix-tree theorem) supplies abundant alternate routes. We
make no claim about rack-to-rack or wide-area distances, where the round-trip
dominates and the analysis differs. Garner's instruction was explicit: state
the assumptions up front; we do.

\paragraph{Three jobs, three mechanisms (no conflation).} The fault model
keeps strictly separate what earlier drafts blurred:

\begin{itemize}[nosep]
\item \textbf{Corruption} is detected by the standard 32-bit FCS, retained
  unchanged. \AE thernet removes the \emph{forward-only} recovery layered on
  top of FCS (RS-FEC, link-level retry), not FCS itself.
\item \textbf{Commitment} is established by the SACK witness chain---which is
  \emph{not} an error-detecting code and does not pretend to be one. A single
  damaged rung fails to witness the next and aborts the transaction cleanly;
  a corrupted SACK is operationally identical to a missing one (a bilateral
  non-commitment), so the eight SACK bits need no ECC. The only dangerous
  event, a false commitment, requires corruption that both escapes the FCS
  \emph{and} forges all four witness rungs consistently---a joint probability
  orders of magnitude below the per-link bit-error rate.
\item \textbf{Link failure} is detected by the Reliable Link-Failure Detector
  (RLFD), which is bilateral (PIF plus a one-hop triangle), not a unilateral
  timeout. Link flapping is its job, not the witness chain's.
\end{itemize}

\paragraph{What ``reversible'' reverses.} Not time, and not retransmission
(\AE thernet does not retry). What reverses is the \emph{commitment state of
the slice engine}, and only before the commit point: when a receiver signals
NAK (``not accepting this,'' \`a la InfiniBand---not a timeout) or a rung
fails to witness, the engine backs out its provisional state to the
pre-exchange state, which each end retains for exactly this purpose. Once the
closer has recorded Commit, nothing reverses it. A single broken link cannot
resolve its own ambiguity (Two Generals). A one-hop neighbour of both ends,
the triangle's \emph{carrier}, resolves it by transporting the one fact that
decides the outcome, whether the closer has recorded Commit; the carrier
never chooses between outcomes. This is two-phase commit specialised to two
parties, with the decision taken by the last participant.
The full state table, the false-commit bound, and how a silent link is
detected from arrivals rather than a timer are in
Appendix~\ref{app:faultmodel}.

\section{Does It Matter? On ``Rarely Occurs in a Good Physical Network''}
\label{sec:cost}

A reasonable objection holds that the faults bilateral commitment guards
against ``only rarely occur in a good physical network,'' and that the right
goal is simply an ever-lower-error link. Two responses.

First, the ``make the link better'' trajectory has stopped paying off
proportionally at AI/HPC scale. At 800~Gb/s optical SerDes the bit-error rate
\emph{per second} has risen, not fallen, as bit rate has scaled, and
correlated link flaps on a $10^5$-GPU cluster occur every minute or two
somewhere in the fabric---documented in \textit{The Ghost in the Datacenter}
\cite{borrill2026ghost} and the Wave critique \cite{borrill2026wave}. The
empirical pattern is more links, faster, with more correlated failure modes,
not fewer faults.

Second, the cost of compensating above the link is no longer small. On
order-of-magnitude estimates, AI-training coordination overhead runs to
\$15--40\,B/yr; cloud-storage synchronization overhead \$5--9\,B/yr; the
high-frequency-trading colocation premium \$1.5--3\,B/yr; the 2024
Delta--CrowdStrike outage cost a single airline on the order of \$500\,M; and
checkpoint compensation consumes an estimated 12--43\% of AI-training time,
depending on cluster size and model architecture. The ``higher layer'' that
absorbs the missing bilateral commitment is, in 2026, an opaque stack of
workarounds---fsync, journaling, end-to-end checksums, consensus rounds,
AllReduce barriers, transaction managers, application retries---whose
aggregate cost now rivals the wire.

Third, and most telling, independent disciplines keep arriving at the same
primitive when they work the problem honestly---which is the real test of
whether ``higher layers handle it'' remains adequate. Wang \emph{et al.}'s
2026 \textit{IEEE TCAD} multi-chiplet result \cite{wang2026inoutbound} reaches,
from a packaging-yield engineering problem with no exposure to this programme,
for an in-place swap of buffered packets between adjacent routers, reserved by
a request and an acknowledgment---the buffer swap of Parasar
\emph{et al.}~\cite{parasar2019swap}---rather than a timeout-and-reserve
recovery, and measures $2.36\times$ higher throughput than the timeout-based
alternative precisely in the over-saturation regime where that alternative
collapses. It is not the bilateral swap of this paper: it exchanges buffered
state rather than committing an outcome, and failure during a swap is outside
its model. But it shows the same pattern: the handshake wins on the very
metric the alternative was meant to optimise, because the alternative was
trying to recover from a problem the handshake prevents. That
result sits on a four-decade trajectory of the same machine at successively
larger scales---bisynchronous FIFO, IEEE~P896 backplane, multi-chiplet NoC,
inter-node link---none using a shared timebase, and none letting a timeout
decide remote state; the
convergence, reviewed in \cite{borrill2026swaptr}, is independent corroboration
of a kind no single-author advocacy can supply. The deeper historical version
of the same point---that the bilateral round-trip was identified as
architecturally necessary, not merely tolerable, by W.~C.~Lynch in 1968 and
restated by him in 2025 as a consequence of signal velocity and the resulting
relativistic situations---is developed in \cite{borrill2026lynchtr}.

This does \emph{not} refute the separation-of-concerns principle Garner
rightly defends: programs will always have bugs, and many problems must be
solved above the network regardless of what the link provides. The claim is
narrower. The bilateral primitive does not \emph{replace} the higher layer;
it \emph{changes the substrate} the higher layer stands on---as virtual
memory changed what operating systems built atop physical RAM. The
higher-layer compensation does not go to zero; it goes to a smaller residual.

\section{The Correction to Table~1}
\label{sec:correction}

\begin{table}[h]
\caption{Two efficiency paradigms.}
\label{tab:comparison}
\centering\small
\begin{tabular}{@{}lll@{}}
\toprule
& \textbf{Metcalfe $E$ (1976)} & \textbf{Bilateral $E_B$ (2026)} \\
\midrule
Measures   & Forward packet throughput      & Committed bilateral transactions \\
Channel    & One-way (forward only)         & Two-way (forward + backward) \\
Success    & ``Packet on Ether, no collision'' & ``Outcome recorded, known at both ends'' \\
Model      & $E=(P/C)/(P/C+W\!\cdot\!T)$    & $E_B=\tfrac{N_c}{N_a}\cdot\tfrac{P_e}{P_e+\Delta T_c}$ \\
Timing     & Slot $T$ (round trip)          & Floor $T\!\ge\!2\tau$; margin $\Delta T_c\!\approx\!0$ \\
Backward   & Invisible to the metric        & First-class channel (PIF / SACK) \\
On failure & Abort \& retry                 & Reversible rollback (triangle) \\
\bottomrule
\end{tabular}
\end{table}

The correction to Table~1 is not a different set of numbers; it is a different
\emph{question}. For fifty years we asked ``how efficiently does the wire
carry packets?'' and celebrated numbers above 98\%. The bilateral
question---``how efficiently does the link produce committed
agreements?''---was never asked, because the assumption that delivery is
purely forward-in-time made it invisible.

\section{Conclusion: Fifty Years On}
\label{sec:conclusion}

The genius of the 1976 paper is that it contains both the forward-throughput
model and the seed of its own complement. The efficiency formula is
forward-only; the end-dally reaches for bilateral agreement; and the deeper
history---Alohanet's positive Acks, removed in 1973---shows the bilateral
mechanism was once \emph{at the link} before being pushed upward. Metcalfe's
NAK remark names the principle fifty years of forward optimization obscured:
systems get more efficient, not less, when backward information flows freely.

Open \AE thernet recovers, at the link and within the rack, the bilateral
commitment that Aloha had and Ethernet dropped. The bilateral transaction is
the primitive; back-to-back Shannon channels with Perfect Information Feedback
are the mechanism; a triangle with reversible rollback is the recovery model.
We honor Metcalfe and Boggs on the fiftieth anniversary not by replacing
Table~1 but by extending it---from forward efficiency to bilateral efficiency,
from packets delivered to transactions committed.

\paragraph{Acknowledgment.} This paper has been improved, substantially and
repeatedly, by Robert Garner (Computer History Museum), whose point-by-point
reviews supplied the SF-10 equation, the PUP/BSP/CSL-79-10 lineage, the 1978
``Ethernet Briefing'' video, the per-file end-dally correction, the
collision-detection reframing toward Alohanet's positive Acks, and the
insistence---adopted here---that the architecture not be called Ethernet and
that ``reversible'' be described as \emph{what and why} rather than as a
slogan. The remaining disagreements are real and are the author's
responsibility.

\bigskip
\noindent\textit{AI Disclosure:} This paper was developed with assistance from
Claude (Anthropic), used as a research and drafting tool. All technical
content, arguments, and conclusions are the author's own.

\bibliographystyle{plainnat}

\clearpage          
\appendix

\ifSubfilesClassLoaded{%
  \newcommand{\fmsec}{\section}%
  \newcommand{\fmsub}{\subsection}%
}{%
  \newcommand{\fmsec}{\subsection}%
  \newcommand{\fmsub}{\subsubsection}%
}

\ifSubfilesClassLoaded{%
  \title{The Open \AE thernet Fault Model:\\
  What the Swap Witnesses, and What ``Reversible'' Reverses}
  \author{Paul Borrill\\
  \textit{D{\AE}D{\AE}LUS}\\
  \texttt{paul@daedaelus.com}}
  \date{v0.2 --- June 10, 2026\\[4pt]
  \small Companion to \textit{The Bilateral Efficiency of Ethernet} (arXiv:2603.19406)}
  \maketitle
}{%
}

\ifSubfilesClassLoaded{%
  \begingroup\renewcommand{\abstractname}{Abstract}%
  \begin{abstract}%
}{%
  \section{The Open \AE thernet Fault Model:
  What the Swap Witnesses, and What ``Reversible'' Reverses}%
  \label{app:faultmodel}%
  \noindent\textit{Synopsis.}\ %
}
This appendix discharges a debt. In March 2026, in response to
Robert Garner's point-by-point review of \textit{The Bilateral Efficiency
of Ethernet}, the author promised ``a clear statement of the \OAE{} fault
model'' as a separate document. This is that document. It answers, at what
Garner rightly called ``an earthly engineering level,'' the three questions
his review left unanswered: (1)~what link fault is the slice-by-slice swap
actually trying to detect, and what value does the swap add over plain
echo-and-compare; (2)~how can a transaction be trusted when there is no
error-correcting code on the eight Slice-ACK (SACK) bits or on the 64-byte
payload; and (3)~what, precisely, is reversed when we speak of ``reversible
state machines,'' since ``you just can't say `reversible state machines'---
you have to say what and why.'' We answer each in turn. The short form: the
32-bit Frame Check Sequence (FCS) is retained and unchanged; what \OAE{}
removes is the \emph{forward-only} recovery machinery (RS-FEC, link-level
retry) layered on top of it. The swap structure is not an error-detecting
code and does not pretend to be one; it is a \emph{witness chain} that
manufactures bilateral mutual knowledge, slice by slice, so that commitment
is an outcome both endpoints reach, each from witnesses it has received,
rather than a guess the sender makes after the fact. ``Reversible'' refers not to time and not to
retransmission but to the \emph{commitment state of the slice engine}: a
provisional commitment that fails to witness is backed out to a symmetric
pre-transaction state that both endpoints agree never happened. This is
Jim~Gray transaction rollback realized in link hardware---not
timeout-and-retry, which \OAE{} eliminates.
\ifSubfilesClassLoaded{%
  \end{abstract}\endgroup
  \tableofcontents
  \bigskip\hrule\bigskip
}{%
  \par\medskip
}

\fmsec{Scope and Assumptions (Stated Up Front)}
\label{sec:scope}

Garner's closing instruction was explicit: \emph{``you need to state the
assumptions up front. If the assumptions are one meter or less, in cellular
networks, you need to be clear about that.''} We comply.

\begin{enumerate}[nosep]
\item \textbf{Distance.} \OAE{} targets links of \textbf{one meter or less}
  (intra-rack), shrinking to sub-centimeter at the chiplet/module boundary.
  At 800~Gb/s a minimum 64-byte frame plus inter-record gap occupies
  $\approx 1.5$~m of wire ($\approx 3$~ns), so within a rack the wire is
  already full with a single frame in flight. This is the regime where the
  fault model below applies. We make \emph{no} claim about rack-to-rack or
  WAN distances; there the round-trip dominates and the analysis is
  different.
\item \textbf{Topology.} A cellular mesh of low-valency nodes (``cells'',
  7--8 ports, octahedral perspective), so that a blocked route has many
  alternative spanning trees (matrix-tree theorem). Failure recovery
  assumes a \textbf{triangle}: two endpoints plus a one-hop third party.
\item \textbf{What is unchanged.} The IEEE~802.3 frame and its 32-bit FCS
  are retained verbatim. \OAE{} is additive at the SerDes/PHY/NIC boundary;
  it does not modify the frame format or the FCS polynomial.
\item \textbf{What this TR does \emph{not} claim.} \OAE{} does not eliminate
  the need for end-to-end verification above the link. A NIC fault, a fault
  on the NIC$\to$CPU path, or an application-level drop is out of scope for
  a link-layer guarantee, exactly as the Saltzer--Reed--Clark end-to-end
  argument requires. \OAE{} improves the foundation; it is not the whole
  building.
\end{enumerate}

\fmsec{Question 1: What Fault Is the Swap Detecting, and Why Not Just Echo?}
\label{sec:q1}

Garner's sharpest objection deserves to be quoted exactly, because the
answer must meet it on its own terms:

\begin{quote}\itshape
``All you're doing is you come up with some random way to scramble the bits
each time. It doesn't provide any extra value unless you believe that the
link itself could have these byte operations on you.'' \\
``Swapping is not adding any value to this, unless you think the link can do
really weird things that only this mechanism could detect.''
\end{quote}

\textbf{He is right, and the paper was wrong to imply otherwise.} The swap
is \emph{not} an error-detecting transform. It does not exist to catch a
link that maliciously permutes bytes. If detecting bit errors were the goal,
plain echo-and-compare plus the FCS would suffice, and the swap would be
ceremony. The confusion is the author's: the slice engine was described as
though its job were detection, when its job is \textbf{commitment}.

\fmsub{The distinction Garner drew is the right one}

Garner: \emph{``A swap operation swaps unrelated things\dots{} this link
thing has got nothing to do with that.''} Correct. There are two senses of
``swap'':

\begin{description}[style=nextline,leftmargin=1.5em]
\item[(a) Semantic swap of unrelated operands.] The SPARC \texttt{SWAP}
  instruction (which Garner helped put in SPARC) exchanges a register and a
  memory word in one indivisible step; the two operands are unrelated; the point is
  deadlock-free mutual exclusion. \OAE{} does \emph{not} do this at the link
  layer, and the paper should never have implied it did.
\item[(b) Structural transform that builds a witness.] What the slice engine
  does is apply a known, invertible transform $T$ (the half-swap $\sigma$) to a slice on the way out
  and verify $T^{-1}$ on the way back. The transform's output is not a more
  reliable copy of the data; it is \emph{evidence that a specific endpoint,
  holding the matching $T^{-1}$, processed this specific slice}. The value
  is epistemic, not error-correcting.
\end{description}

\fmsub{What echo-and-compare cannot give you}

Echo-and-compare gives the \emph{sender} a one-way, after-the-fact check:
``the bits I got back match the bits I sent, so probably they arrived.''
Three things it does not give:

\begin{enumerate}[nosep]
\item \textbf{Knowledge at both ends.} After echo-and-compare the
  \emph{receiver} knows nothing it can act on: it echoed bytes without
  committing to them. Only the sender learns anything, and only about the
  past. \OAE{}'s exchange makes each step a \emph{joint} one: the receiver
  advances its own state only by emitting the witness that lets the sender
  advance, and vice versa. The knowledge reached is nested to finite depth
  and asymmetric. When Commit arrives, the closer knows that the initiator
  verified delivery; when Done arrives, the initiator knows that the closer
  has committed; the closer cannot know that Done arrived. That is not
  common knowledge, which no finite exchange over an unreliable channel
  attains \cite{halpern1990knowledge}, and the outcome does not need it: the
  custody rule of Section~\ref{sec:q3} fixes it. Bill Lynch's
  alternating-bit protocol was reaching for the same per-message knowledge,
  without quantum language.
\item \textbf{Slice-by-slice commitment.} Echo-and-compare is whole-frame:
  you learn the outcome after the frame round-trips. The witness chain
  commits progressively (8\,$\to$\,16\,$\to$\,32\,$\to$\,64 bytes), so a
  failure is localized to the rung at which the witness broke, and the
  un-commitment (Section~\ref{sec:q3}) is correspondingly partial and clean.
\item \textbf{A backout point.} Echo-and-compare has no notion of
  \emph{un}-sending. The witness chain does: each rung is a checkpoint the
  slice engine can roll back to.
\end{enumerate}

\fmsub{What the swap genuinely does \emph{not} help with (conceded)}

Garner: \emph{``swapping ain't going to help''} burst errors, and he is
right. Burst errors are caught by the FCS, full stop. The swap chain is
orthogonal to error detection. We state this plainly so the claim is not
oversold: \textbf{the FCS detects corruption; the witness chain establishes
commitment; link flapping needs a third mechanism entirely}
(Section~\ref{sec:rlfd}). Three jobs, three mechanisms, no conflation.

\fmsec{Question 2: No ECC on 8 SACK Bits or 64 Bytes---Why Trust It?}
\label{sec:q2}

\fmsub{The FCS is still there}

The most common misreading---and the paper invited it---is that \OAE{}
discards the integrity machinery. It does not. The IEEE~802.3 32-bit FCS is
present and unchanged. What \OAE{} removes is the \emph{forward-only}
recovery layer that conventional fabrics stack \emph{on top of} the FCS:
Reed--Solomon FEC, link-level retry buffers, NAK/timeout retransmission
state. Perfect Information Feedback (PIF) sits on top of the FCS, not in
place of it. If a payload is corrupted in flight, the FCS catches it; the
terminal witness (SACK~11) does not form because the frame does not match a
legitimate layout; the transaction aborts with both ends knowing why.

\fmsub{Eight SACK bits are too few for ECC---and do not need it}

Garner is correct that eight bits cannot carry a useful error-correcting
code. The witness chain makes ECC on the SACK bits unnecessary, for the same
structural reason that the chain replaces forward FEC: \textbf{each rung is
only meaningful as a witness to the prior rung.}

\begin{equation}
\label{eq:chain}
\text{SACK 00} \;\rightarrow\; \text{SACK 01} \;\rightarrow\;
\text{SACK 10} \;\rightarrow\; \text{SACK 11}
\end{equation}

SACK~11 is accepted only if SACK~10 was accepted, which required SACK~01,
which required SACK~00. A single-bit error on any rung produces a witness
that does not match the expected $T^{-1}$ pattern at the \emph{next} rung;
the swap halts at that rung; the transaction aborts cleanly; both ends know
\emph{which} rung failed. \textbf{A corrupted SACK is operationally
identical to a missing SACK: a bilateral non-commitment.} There is nothing
to ``correct,'' because the protocol never needs to recover a SACK---it
needs only to know whether bilateral commitment was reached, and a damaged
rung answers ``no'' as definitively as a clean one.

\fmsub{The false-positive bound}

The only failure that matters is a \emph{false} SACK~11: the protocol
declaring commitment when it did not occur. This requires the conjunction of
two independent rare events:

\begin{equation}
\label{eq:fp}
P_{\text{false commit}}
\;\le\;
\underbrace{P_{\text{FCS escape}}}_{\approx 2^{-32}\ \text{for a CRC-32 burst}}
\times\;
\underbrace{P_{\text{rung-pattern match} \mid \text{corruption}}}_{\text{all four rungs corrupt consistently}}
\end{equation}

For the terminal commitment to be falsely accepted, the corruption must
(i)~escape the 32-bit FCS \emph{and} (ii)~corrupt all four SACK rungs into a
sequence that matches the expected witness pattern for a legitimate frame
layout. These are independent; their joint probability sits orders of
magnitude below the per-link bit-error rate. This is what is meant by ``the
bilateral structure of the swap \emph{is} the integrity mechanism'': not
that the swap adds redundancy, but that commitment is gated behind a
conjunction of conditions that a corrupting link cannot satisfy by accident.

\textbf{What this is not.} It is not a claim of zero error. It is a claim
that the residual false-commit rate is bounded by
Equation~\eqref{eq:fp} and is therefore engineerable to any target by
choice of frame layout and witness depth---and that the residual is
\emph{reported}, because a transaction that cannot establish bilateral
commitment fails closed (it does not commit) rather than failing silent (the
mode the 1975 MAXC/station-201 corruption stories warn against).

\fmsec{Question 3: What Does ``Reversible'' Reverse? (What and Why)}
\label{sec:q3}

Garner's instruction was the most useful sentence in the meeting:

\begin{quote}\itshape
``You would want to more carefully describe what's going backwards. What
does it mean for the state machine? You just can't say `reversible state
machines.' You have to say, you know, what and why.''
\end{quote}

So: \textbf{what} and \textbf{why}.

\fmsub{What is \emph{not} reversed}

\begin{itemize}[nosep]
\item \textbf{Not time.} Nothing in \OAE{} runs time backward. The hardware,
  the observers, and the signals all proceed forward, exactly as Garner
  insists. ``Everything is forward in time'' is granted without reservation
  for this document.
\item \textbf{Not retransmission.} \OAE{} does not retry. There is no
  retransmission buffer, no resend-on-timeout, no ``keep trying until it
  works.'' Reversal is the \emph{opposite} of retransmission: instead of
  re-driving the wire forward, the slice engine \emph{withdraws} a
  commitment it had provisionally advanced.
\end{itemize}

\fmsub{What \emph{is} reversed: the commitment state of the slice engine}

The slice engine is a finite state machine whose states are the rungs of the
witness chain. ``Reversible'' means each forward transition has a defined
inverse transition that returns the engine to the prior rung \emph{and}
emits the signal that lets the peer do the same---so that after a reversal
\textbf{both endpoints agree the advance never happened}. The unit of
reversal is a \emph{provisional commitment}, not a packet and not a clock
tick.

\begin{table}[h]
\centering
\small
\caption{Slice-engine states and the reverse transitions. The forward
transition advances a provisional commitment; the reverse transition backs
it out symmetrically. NAK here means ``not accepting this'' (\`a la
InfiniBand), \emph{not} a timeout. The four SACK rungs (00--11) are
named, respectively, SHANNON-DALLY, SPEKKENS-DALLY, TURING-DALLY, and
METCALFE-DALLY; for a fuller description of each rung and the figure whose
work it anchors, see \cite{borrill2026amw52}.}
\label{tab:states}
\begin{tabular}{@{}lll>{\raggedright\arraybackslash}p{4.4cm}@{}}
\toprule
\textbf{State} & \textbf{Forward in} & \textbf{Reverse on} & \textbf{What the reversal restores} \\
\midrule
\textsc{idle}    & first slice arrives & --- & (nothing committed yet) \\
SACK 00          & 8\,B witnessed      & NAK / no-witness & link returns to \textsc{idle}; no slice retained \\
SACK 01          & 16\,B witnessed     & NAK / no-witness & back to SACK 00; 16\,B capture discarded \\
SACK 10          & 32\,B witnessed     & NAK / no-witness & back to SACK 01; partial decode discarded \\
SACK 11          & 64\,B witnessed, FCS valid & (terminal; commit) & --- \\
\bottomrule
\end{tabular}
\end{table}

\fmsub{Why reversal is required: a single broken link is unrecoverable
from two parties}

This is the part the paper never argued, and it is the load-bearing one.
Garner: \emph{``they're dead---who cares whether they reverse?''} The answer
comes from Jim Gray's contract theory, by way of the Mulligan Stew Gang. (The full
genealogy---that Gray's 1981 atomicity is, on our reading, \emph{bilateral mutual
commitment} (the marriage ceremony's ``I do''), and that the field
industrialized only its forward-time logging face---is the subject of a
companion report in the Category Mistake in Networking series
\cite{borrill2026graytr2007}.)

A two-party commitment over a single link has no safe outcome \emph{from
the link alone} when the link dies mid-exchange: the sender cannot tell
``received and committed'' from ``lost on the way back,'' and no amount of
waiting on that link resolves it---this is the Two Generals impossibility,
and a timeout merely converts ``I~don't know'' into ``I~gave up.''

Gray explained two-phase commit with the marriage ceremony: the
participants say ``I~do,'' and a participant that has said it gives up the
right to abort unilaterally. On our reading, the \OAE{} exchange carries
that ceremony onto the link with no minister in the steady state. The
initiator says ``I~do'' when it sends Commit; the closer pronounces by
recording Commit recoverably, and that recording is the commit point. The
outcome is therefore fixed by one fact held at one end. The \textbf{custody
rule}: the closer's port processes the arrival of Commit and its own
declaration of failure in a single order, so its state freezes either
committed or uncommitted, and a Commit arriving after the freeze is
discarded. The \textbf{triangle} supplies what the broken link cannot: any
node adjacent to both ends can act as a \emph{carrier}, learning the closer's
frozen state over its own edge and delivering it to the initiator. The
carrier never chooses between outcomes, so another common neighbour can
replace it without changing the result.

Reversibility is what makes the back-walk well-defined \emph{before} the
commit point. If the closer's state froze uncommitted, each endpoint's slice
engine walks \emph{back} down Table~\ref{tab:states} to \textsc{idle},
restoring the pre-exchange slot it retained, and the two ends re-converge on
``nothing committed.'' After the commit point nothing walks back. Retention
and the custody rule, not invertibility alone, are what keep one end from
staying committed while the other is not, and the guarantee is that both
ends \emph{reach} the same outcome, not that they hold it at every instant:
between Commit being sent and Done landing, the closer may have committed
while the initiator is in doubt. Two cases still block, as two-phase commit
does: the closer failing between the commit point and the delivery of the
outcome, and a partition that separates the ends from every common
neighbour \cite{borrill2026hospital}. \textbf{Reversal is rollback before
the commit point; the closer decides; the carrier transports the decision;
the link engine is the participant.} That is the what and the why.

\fmsub{No timer: how a silent link becomes an event}
\label{sec:notimer}

Garner caught a real thing: the word ``timeout'' appears in the
SystemVerilog of the \OAE{} Specification v0.8, as a counter of
fabric-clock cycles that bounds how long the slice engine waits for a
witness. Earlier drafts defended it as \textbf{timeout-as-trigger} (it may
start an investigation) rather than \textbf{timeout-as-truth} (it may not
decide remote state), the distinction the Wang \emph{et~al.}
In/Outbound-Swap NoC mechanism also observes \cite{wang2026inoutbound}.
That distinction stands. What it conceded was more than the architecture
needs: the trigger does not have to come from a clock.

A link appears to need a timer because a single endpoint, alone with one
link, has only silence to go on when its peer stops. Absence produces no
event, so something local must eventually manufacture one. An \OAE{} cell
is not alone with one link. It is octavalent, and an agent at each cell
owns its eight ports and keeps tokens circulating, clockwise and
anticlockwise, around the one-hop triangles it belongs to
\cite{borrill2026hospital}. Each token returns to the agent as an arrival:
either completed, or reversed by the neighbour that could not forward it,
carrying that neighbour's four-state view of its own edges. A token is
re-emitted on its own return, so the loops are self-timed and need no
period set by a clock. The agent therefore receives a continuous stream of
events from its other ports, and a silent link is detected against that
stream. Three sources of event suffice, and none of them counts time:

\begin{enumerate}[nosep]
\item \emph{Reversal.} A neighbour that cannot forward a token across the
  suspect edge returns it with its own report. Silence on one edge becomes
  an arrival on another.
\item \emph{Lapping.} If the witness on edge $e$ has not returned while the
  loops through the agent's other edges have each completed $\Xi$ returns,
  the agent marks $e$ suspect. The counter that does this is incremented by
  the arrival of sibling tokens, not by an oscillator. It has no frequency
  and no drift, and it scales with the cell: if the whole cell slows (a
  stall, a pause, thermal throttling), its sibling loops slow with it and
  nothing is falsely suspected. That false suspicion of the merely slow is
  the failure mode that makes ordinary timeouts unreliable.
\item \emph{Signal state.} A SerDes peer that dies stops driving the line;
  the receiver's loss-of-signal or loss-of-lock indication is a change of
  physical state, delivered as an event.
\end{enumerate}

\noindent Suspicion produced this way is still only a trigger. It starts
custody (Section~\ref{sec:rlfd}); the outcome of any transaction in flight
is then fixed by a fact held at one endpoint and carried by a third node,
never by the suspicion itself.

What remains is an assumption, and it is stated rather than hidden.
Lapping requires that while the sibling loops complete $\Xi$ cycles, a live
loop through $e$ completes at least one. That bounds the \emph{ratio} of
the cycle times of causal loops that share a node; it is not a bound in
seconds. The idea that such ratios can stand in for time is not ours. The
Asynchronous Bounded-Cycle model of Robinson and Schmid is entirely
time-free: it constrains only the relative lengths of certain causal chains
of messages, and is strong enough to simulate lock-step rounds and solve
consensus \cite{robinson2008abc}. The $\Theta$-model, which bounds the
ratio of the longest to the shortest message delay in transit at the same
time, supports a perfect failure detector and atomic commitment without
clocks \cite{widder2004theta}; and Most\'efaoui, Mourgaya and Raynal build
failure detectors from the order in which responses arrive rather than from
timers \cite{mostefaoui2003async}. What is offered here is the application
of that idea at the granularity of one link, fed by counter-rotating
tokens. On a substrate that violates the ratio bound (a live loop starved
indefinitely while its siblings run), slow and failed remain
indistinguishable and no protocol changes that; the impossibility results
are not evaded. A cell that loses all its ports at once has no sibling
stream: it is partitioned, and falls under CAP. A two-port repeater cell
belongs to no triangle and has only reversal from its neighbours and signal
state.

The claim is therefore narrow and checkable. \OAE{} uses no synchronised
clock anywhere, and no local timer on the failure path: every failure event
is an arrival. The cycle counter Garner read in the v0.8 SystemVerilog is
to be replaced by a lap counter driven by sibling-token arrivals; until the
specification makes that change, the counter is a placeholder and should be
read as one.

\fmsec{The Reliable Link-Failure Detector (RLFD)}
\label{sec:rlfd}

The triangle is also what upgrades failure detection from \emph{unreliable}
to \emph{reliable}. A single link cannot reliably detect its own failure
(the impossibility above). With the triangle, the \OAE{} \textbf{Reliable
Link-Failure Detector (RLFD)} is bilateral: PIF gives committed-transfer
liveness on each edge, Triangle Failover gives an independent witness to a
suspected-dead edge, and Atomic Token Transfer gives a single,
agreed-upon owner of the recovery (the outcome of an exchange in flight is
still fixed by the closer's custody rule, not by that owner). The events that drive
all three are harvested by the octavalent agent of
Section~\ref{sec:notimer}, so the detector has no timer of its own. The ``Reliable'' is earned from
the bilateral physics of the link plus the triangle, not from a unilateral
timeout heuristic. This is explicitly \emph{not} a Chandra--Toueg
``unreliable failure detector,'' which presumes the very
timeout-and-retry \OAE{} removes. Link flapping---the burst-correlated,
``every minute or two on a $10^5$-GPU cluster'' failure mode---is the RLFD's
job, not the witness chain's; see \textit{The Ghost in the Datacenter}
\cite{borrill2026ghost}.

\fmsec{Summary: Three Jobs, Three Mechanisms, No Conflation}
\label{sec:summary}

\begin{table}[h]
\centering\small
\caption{The separation Garner asked for, made explicit.}
\begin{tabular}{@{}lll@{}}
\toprule
\textbf{Job} & \textbf{Mechanism} & \textbf{Not done by} \\
\midrule
Detect corruption        & 32-bit FCS (unchanged)      & the swap \\
Establish commitment     & witness chain (SACK 00--11) & the FCS \\
Detect link failure      & RLFD (triangle + PIF + agent) & a timer \\
Recover from failure     & rollback before commit; carrier & retransmission \\
\bottomrule
\end{tabular}
\end{table}

The swap was never an error-detecting code; saying so was the author's
imprecision and Garner's correction is adopted. The eight SACK bits need no
ECC because a damaged witness is a non-commitment, and the false-commit rate
is bounded by the conjunction in Equation~\eqref{eq:fp}. ``Reversible''
reverses the slice engine's \emph{commitment state}---Gray-style rollback in
hardware, gated by a triangle---never time and never retransmission. Stated
this way, at the engineering level Garner asked for, the mechanism either
stands or can be falsified on its own terms. That was the goal.

\bigskip
\noindent\textit{AI Disclosure:} This report was developed with assistance
from Claude (Anthropic), used as a research and drafting tool. All technical
content, arguments, and conclusions are the author's own.

\ifSubfilesClassLoaded{%
\bibliographystyle{plainnat}

}{}

\end{document}

\clearpage
\section{The Acknowledgment Returns: Link-Level Retry as Half an End-Dally}
\label{app:ackreturns}

In September 2026 Metcalfe recalled for the author how Ethernet first
became ten times faster. The observation, which he credits mostly to the
late Ron Crane, was that most Ethernets were not a kilometer long but lay
within about 100 meters, so the same arithmetic worked at ten times the
speed; the result was 100~Mb/s Fast Ethernet \cite{metcalfe2026crane}. That arithmetic is the arithmetic
behind this paper's scope. This appendix carries it forward to 100~Gb/s,
and then records what the industry is doing at the end of it.

\subsection{Crane's arithmetic, carried to 100~Gb/s}

The fixed quantity in Ethernet is the minimum frame: 64~bytes, or 512~bits,
from destination address through FCS. Under CSMA/CD it was also the slot
time, the shortest transmission that still let a station hear a collision
from the far end, so the network's round trip had to fit inside it. A frame
of fixed size gets shorter in space as the bit rate rises. With signals
travelling at about $2\times10^{8}$~m/s (roughly 5~ns per meter in copper
twinax or in fiber):

\begin{center}\small
\begin{tabular}{@{}rrrl@{}}
\toprule
\textbf{Rate} & \textbf{512 bits take} & \textbf{Length in space} & \textbf{What happened} \\
\midrule
10~Mb/s  & 51.2~$\mu$s & $\sim$10~km & 2.5~km maximum network diameter \\
100~Mb/s & 5.12~$\mu$s & $\sim$1~km  & $\sim$200~m collision domain (Fast Ethernet) \\
1~Gb/s   & 512~ns      & $\sim$100~m & half duplex needed carrier extension \\
         &             &             & to a 4096-bit slot \\
10~Gb/s  & 51.2~ns     & $\sim$10~m  & full duplex only; CSMA/CD dropped \\
100~Gb/s & 5.12~ns     & $\sim$1~m   & the scope of this paper (Section~\ref{sec:scope-note}) \\
800~Gb/s & 0.64~ns     & $\sim$13~cm & sub-meter links of Section~\ref{sec:shannon} \\
\bottomrule
\end{tabular}
\end{center}

Crane's observation was that the frame had shrunk to fit the networks
people were actually building. The trick worked once more. At 1~Gb/s the
shared medium ran out of room, and from 10~Gb/s on Ethernet gave up the
collision domain and became point-to-point and full duplex. The constraint
did not go away; it stopped being enforced, because nothing on a
full-duplex link needed it.

At 100~Gb/s the minimum frame is about one meter long, which is exactly
where this paper's scope stops. A one-meter link has a round trip of about
10~ns. A minimum frame occupies 5.12~ns of MAC time, or 6.72~ns once the
8-byte preamble and SFD and the 12-byte inter-frame gap are counted (148.8
million minimum frames per second). The forward and backward paths of a
one-meter link therefore hold about one and a half minimum frames between
them. That is the regime of Metcalfe's ``the snake fills the building''
(Section~\ref{sec:shannon}): an acknowledgment returning behind a frame spends
no forward bandwidth, because nothing else could have used the backward
path.

\paragraph{The lane arithmetic.} None of this depends on how many lanes
carry the 100~Gb/s, and the point is easy to get wrong. IEEE~802.3 defines
100~Gb/s Ethernet over
\begin{itemize}[nosep]
\item four lanes (100GBASE-CR4/KR4/SR4): $4\times25.78125$~Gb/s NRZ with
  RS(528,514) FEC, 103.125~Gb/s on the line;
\item two lanes (100GBASE-CR2/KR2/SR2): $2\times53.125$~Gb/s PAM4
  (26.5625~GBd), RS(544,514), 106.25~Gb/s;
\item one lane (100GBASE-CR1/KR1/DR): $1\times106.25$~Gb/s PAM4 at
  53.125~GBd with RS(544,514).
\end{itemize}
There is no three-lane 100~Gb/s PHY. The PCS stripes every frame across
all the lanes, so on $L$ lanes each lane carries $512/L$ of the frame's
bits at $1/L$ of the rate, and the frame still occupies 5.12~ns. It does
not take $4\times5.12=20.48$~ns on a four-lane link; that is the time a
frame would take on a single 25~Gb/s lane, and 100GBASE-R never sends one
that way. Lane count changes the baud rate and the modulation, not the
length of a frame in time or in space.

The FEC codeword is invariant in the same way. RS(528,514) at 103.125~Gb/s
and RS(544,514) at 106.25~Gb/s both last exactly 51.2~ns, and both carry
5{,}120 MAC bits: ten 512-bit slots, or about 7.6 minimum frames once
preamble and gap are counted. (These are the digits of the 10~Mb/s slot
time, a thousand times faster.) This matters for retry. The unit the PHY
loses is an uncorrectable codeword, not a frame, so one bad codeword can
take out as many as eight minimum frames. A one-meter link's 10~ns round
trip is a fifth of one codeword. In practice the loop is dominated by
SerDes, DSP and FEC-decode latency rather than by flight time, as an
operator's contribution to IEEE~802.3 notes \cite{hu2026plsg}. The claim
here is that propagation no longer stands in the acknowledgment's way, not
that the acknowledgment is free.

Read forward, Crane's observation says that a link short enough for the
minimum frame to fill it is a link where the acknowledgment Metcalfe
removed in 1973 can come back. That is what the industry is now doing.

\subsection{The fifth swing}

Section~\ref{sec:zigzag} traced four swings between bilateral and
unilateral mechanisms, from Alohanet's positive Acks to TCP. Between 2025
and 2026 a fifth swing began, and it moves back toward the link. It bears
directly on the paper's claim: the industry is restoring at the link part
of what Metcalfe removed in 1973. The question is how much of it.

The acknowledgment has returned to Ethernet links through industry consortia
rather than through IEEE~802. The Ultra Ethernet Consortium's specification
(release 1.0.3, 16 July 2026) defines link-level retry (LLR), in
which the receiver acknowledges or negatively acknowledges frames and the
sender holds a replay buffer, and credit-based flow control (CBFC)
\cite{uec2026spec}. Release 1.1 (12 October 2026) extends the specification
to scale-up networks, with a profile for the short accelerator-to-accelerator
links inside a node \cite{uec2026v11,emmons2026nww}: the links of
Section~\ref{sec:scope-note}. The Open Compute Project's Ethernet for Scale-Up
Networking specification, ESUN~1.0 (released 10 March 2026), adopts UEC's
LLR and CBFC rather than defining its own acknowledgment protocol, and
replaces the IP/UDP stack with a four-byte header \cite{ocp2026esun}. UALink,
which never presented itself as lossy Ethernet, carries link-level
retransmission and credit-based flow control from the start
\cite{ualink2025spec}. LLR and CBFC were first shown interoperating between
independent implementations at full 800GE line rate at OFC in March 2026
\cite{keysight2026ofc}.

IEEE~802 is supplying the encoding afterwards. IEEE~802.3 chartered an
Ethernet Metadata Services study group on 1 August 2025, which became the
P802.3dt task force on PAR approval on 12 February 2026 \cite{ieee8023dt}.
IEEE~802.1's own alternative to pause-based flow control, P802.1Qdw Source
Flow Control, has had an approved PAR since 21 September 2022; when its project page was
read on 18 September 2026 it listed no available draft and a vacant editor
position, and the PAR is valid only to 31 December 2026 \cite{ieee8021qdw}.

Operators are now bringing measurements into both working groups. On
17 September 2026 China Mobile asked IEEE~802.1 Nendica whether and how
802.1 should take up link-layer retransmission, describing LLR as recovering
FEC-uncorrectable loss hop by hop ``before it surfaces to upper layers as
timeouts and retries'' \cite{hu2026nendica}. Its measurements, from an
FPGA-emulated single-hop testbed doing memory-semantics writes
\cite{liu2026sigcomm}, put the acknowledgment traffic at about 1\% of
bandwidth with baseline goodput unchanged, and the goodput loss at a 2\%
injected packet-error rate at about 13\% with LLR against about 43\%
without. The same group had already taken LLR to the IEEE~802.3 400~Gb/s
per-lane study group, arguing by analogy with forward error correction,
which went from optional in 802.3ap to mandatory in 802.3bj: when analog
margin alone is no longer enough, a digital recovery mechanism is added
\cite{hu2026plsg}. These are one operator's results on one hop, not a
survey; but they are the cost argument of Section~\ref{sec:bilateral}, made
with someone else's numbers in IEEE's own rooms.

This is consistent with Garner's account in Section~\ref{sec:intro}. Removing
the link acknowledgment was the right decision for the links of 1973; the
consortia are restoring it for sub-meter links at 800~Gb/s, where the
argument of Section~\ref{sec:zigzag} predicts that dropping it becomes
expensive.

\subsection{Four statements, three of them messages}

The word ``acknowledge'' is now carrying four different statements, and only
three of them are messages exchanged between the endpoints.

\begin{center}\small
\begin{tabular}{@{}lll@{}}
\toprule
\textbf{Statement} & \textbf{Content} & \textbf{Kind} \\
\midrule
CREDIT  & You may use $N$ units of my buffer & permission \\
ACK     & I received everything through $N$ & knowledge of an event \\
NAK     & $N$ is missing, or refused for reason $r$ & contradiction \\
TIMEOUT & I have not seen the evidence I expected & inference from absence \\
\bottomrule
\end{tabular}
\end{center}

\noindent
The first three are positive information sent from one endpoint to the
other. The fourth is a conclusion one endpoint draws on its own: a timer
expiring at the sender shows only that the sender has not seen an
acknowledgment by time~$T$. It does not show that the receiver lacks the
frame; the receiver may hold it and the acknowledgment may have been lost or
delayed. LLR's NAK is Metcalfe's observation from Section~\ref{sec:end-dally}
built into a link layer: backward information lets the system converge
sooner than waiting does. Every deployed mechanism listed below narrows the
role of the timer. None of them removes it. UEC's own designers bound the
link-level retransmission timeout by the link round trip, about a
microsecond \cite{hoefler2025ue}, and Broadcom's overview of UEC LLR lists
timer-based retransmission as the protection against lost NACKs and against
a frame that never arrives \cite{opsasnick2026itu}. The proposed row replaces the
timer with arrivals, under a stated ratio assumption
(Appendix~\ref{sec:notimer}).

\subsection{Link-level retry is half an end-dally}

Compare LLR with the EFTP end-dally of Section~\ref{sec:end-dally}, which has
three phases.
\begin{enumerate}[nosep]
\item The sender transmits END.
\item The receiver answers ENDREPLY and dallies.
\item The sender echoes ENDREPLY. Only now is the receiver ``assured''.
\end{enumerate}
LLR has the first two phases and not the third. A frame goes forward; an ACK
or NAK comes back; on ACK the sender releases its replay buffer. Write
$K_S(\phi)$ for ``the sender knows $\phi$'' and $K_R(\phi)$ for ``the receiver
knows $\phi$'', with $\phi$ the statement that frame~$N$ was delivered intact.
After a successful LLR exchange $K_R(\phi)$ and $K_S(\phi)$ hold, but
$K_R K_S(\phi)$ does not: the receiver never learns that its ACK arrived. In
the paper's terms the sender is assured and the receiver is not. When an ACK
is lost, the gap is closed by the sender's replay timer: the sender times
out, replays, and the receiver discards the duplicate and acknowledges again.
The protocol delivers correctly, but the ambiguity is resolved by the
inference in the last row of the table rather than by a message.

This is the precise sense in which link-level retry is half a bilateral
commitment. It restores delivery agreement at the link, which Ethernet gave
up in 1973, but not mutual knowledge of that agreement, which the end-dally
was reaching for. Under the definition of Equation~\eqref{eq:E-bilateral},
an LLR-acknowledged frame is delivered but not committed, because committed
requires that both sides know the outcome. LLR therefore raises forward
goodput on a lossy link and leaves $E_B$ where it was.

The four-message exchange of \OAE\ supplies the third phase. Commit reaches
the closer only after the initiator has verified the returned slice, so
$K_R K_S(\phi)$ holds there, and Done tells the initiator that the closer has
committed. It does not reach common knowledge either, and no finite exchange
over an unreliable channel can \cite{halpern1990knowledge}: the closer cannot
know that Done arrived. The outcome does not depend on that last step; it is
fixed by a custody rule (Appendix~\ref{sec:q3}).

\subsection{Where each mechanism stops}

\begin{center}\footnotesize
\setlength{\tabcolsep}{3.5pt}
\begin{tabular}{@{}lllll@{}}
\toprule
\textbf{Mechanism} & \textbf{Feedback} & \textbf{Sender learns} & \textbf{Receiver learns} & \textbf{Timer role} \\
 & & & \textbf{that sender knows} & \\
\midrule
PCIe data link layer & ACK/NAK DLLP & delivery & --- & replay timer \\
InfiniBand RC & ACK/NAK & delivery & --- & retry timer \\
UEC LLR & ACK/NAK at the link & delivery & --- & backstop \\
UEC CBFC & credit updates & buffer room & --- & --- \\
OCP ESUN & UEC LLR + CBFC & delivery, room & --- & inherited \\
UALink & link retry + credits & delivery, room & --- & replay timer \\
\midrule
\OAE\ (proposed) & bilateral slot outcome & outcome & yes & none (arrivals) \\
\bottomrule
\end{tabular}
\end{center}

\noindent
The column that matters is the fourth. In every deployed mechanism the
receiver knows what it received, and the sender learns of it; in none does the
receiver learn that the sender knows. The last row is a proposal rather than a deployed
mechanism, and its fault model is Appendix~\ref{app:faultmodel}. It is
offered not as a better way to add acknowledgments to Ethernet, where the
consortia arrived first, but as a change in what is acknowledged: the shared
state transition rather than an event at a layer.

\subsection{What retry does with a sick link}

Link-level retry repairs a loss; it does not classify a link. A link that
corrupts intermittently is mended frame by frame: each replay succeeds, the
link stays in service, and the damage appears as a latency tail rather than
as a fault. That is the correlated, flapping failure that dominates large
clusters \cite{borrill2026ghost}, and it is the question Garner has pressed
hardest about this proposal: what does a detector that can only declare
failure do with a link that is sick rather than dead? The deployed
mechanisms do not have to answer, because their retry hides the symptom.

The \OAE\ answer is Link Hospital \cite{borrill2026hospital}. A link
reported faulty from both ends is held: both ends freeze their swap state,
which each end already retains. The octavalent agent then
probes it by sending both rotations of a triangle through the patient, and
readmits it when both come back; repeated probe failure retires it. Every
transition is triggered by an arrival, not by a timer on the patient. How
many consecutive clean periods readmission needs is policy, and we do not
claim to have derived it; that a period is counted in arrivals rather than
seconds is not policy. A link-agnostic definition of a flap, written as a
proposed Open Compute contribution, is in OAE-TR-0096
\cite{borrill2026flap}.

\subsection{Both ends of the stack}

Restoring knowledge at the link pays only if something above it uses that
knowledge. Two companion papers presented at Netdev~0x1A carry the argument
to the two ends of the Linux networking stack. At the southbound end, toward
the wire, \emph{Promise Networks} \cite{singhaijain2026promise} pushes the
end-dally down into the link as an admission primitive: every frame is gated
on a token the peer has issued, so a sender without one cannot transmit.
Congestion control then becomes admission control inside the network rather
than a feedback loop above it, which interprets loss, delay and ECN marks
after the fact. UEC's CBFC is the nearest deployed relative, and the
difference is the one in the table above: a credit grants buffer space, while
a promise commits both ends to an outcome. At the northbound end, where the
stack meets the application, \emph{Why Syscalls Fail}
\cite{singhai2026syscalls} finds the same missing third phase at the
kernel/userspace boundary: most system calls offer at most one-sided
confirmation, so a failure between the kernel and the application is
usually silent. The rule that joins the layers is the same at every
boundary: a layer may report an exchange as done to the layer above only
when its own outcome is recorded and the layer below has reported the
exchange complete, not merely delivered. An end-dally restored only at the link leaves both ends of
the stack unilateral; those two papers are where the rest of the work is
proposed.

\paragraph{Scope of this appendix.} The release dates and project status
above were checked against the issuing bodies' own pages; the mechanism
descriptions are given only at the level of detail on which the primary
sources and the published descriptions agree. A fuller
survey of the 2025--2026 mechanisms, including endpoint-side selective
retransmission schemes and the history of credit-based flow control, with
each claim graded and conflicts between sources recorded, is in
OAE-TR-0050 \cite{borrill2026acktr}; claims that report graded below
``verified'' are deliberately left out here.

\end{document}